# Minding non-collapsibility of odds-ratios when recalibrating risk prediction models

**Mohsen Sadatsafavi**[1], **Hamid Tavakoli**[1], **Abdollah Safari**[1,2]

1. Respiratory Evaluation Sciences Program, Collaboration for Outcomes Research and Evaluation, Faculty of Pharmaceutical Sciences, The University of British Columbia

2. Department of Mathematics, Statistics, and Computer Science, University of Tehran

**Corresponding Author:**
>Mohsen Sadatsafavi, PhD
>Associate Professor, Faculty of Pharmaceutical Sciences
>The University of British Columbia
>+1 604 827 3020
>http://resp.core.ubc.ca/team/msafavi
>msafavi@mail.ubc.ca

**Word count:** 1,552

**Abstract word count:** 150

**Conflict of interest:** None declared.

**Source of Funding:** Financial support for this study was provided entirely by the Canadian Institutes of Health Research (CIHR). The funding agreement ensured the authors' independence in designing the study, interpreting the data, writing, and publishing the report.

**Abstract**

In clinical prediction modeling, model updating refers to the practice of modifying a prediction model before it is used in a new setting. In the context of logistic regression for a binary outcome, one of the simplest updating methods is a fixed odds-ratio transformation of predicted risks to improve calibration-in-the-large. Previous authors have proposed equations for calculating this odds-ratio based on the discrepancy between the prevalence in the original and the new population, or between the average of predicted and observed risks. We show that this method fails to consider the non-collapsibility of odds-ratio. Consequently, it under-corrects predicted risks, especially when predicted risks are more dispersed (i.e., for models with good discrimination). We suggest an approximate equation for recovering the conditional odds-ratio from the mean and variance of predicted risks. Brief simulations and a case study show that this approach reduces under-correction, sometimes substantially. R code for implementation is provided.

**Background**

When transporting a prediction model from the population in which it is developed to a new one, predictions might be systematically under- or over-estimated. Thus, prediction models often require recalibration in the new population, or periodically over time in the same setting – a practice referred to as model updating[1]. One of the simplest updating methods is to adjust the 'intercept' of the model, without changing predictor effects, to improve the model's calibration-in-the-large (the difference between average predicted and observed risks)[2]. In the context of logistic regression for binary outcomes, this is equivalent to applying a fixed odds-ratio to predicted risks. We are aware of two related contexts where this approach has been advocated. The first is when the model is updated to account for differences in outcome prevalence between the original and target populations[2]. For example, if a model is developed in a population with outcome risk of 25% (odds of 1/3), and is now applied to a population with an outcome risk of 20% (odds of 1/4), one would apply the odds-ratio of (1/4)/(1/3)=3/4 to predicted risks. This method assumes that the change in prevalence is independent of included predictors. The second context is when we have a sample from a new target population. Janssen et al applied a prediction model for post-operative pain, developed in a sample with an average risk of 62.0%, to an independent sample with an average risk of 36.1%[3]. The average predicted risk in this sample was 57.7%. They proposed changing the intercept of the model by a factor corresponding to an odds-ratio of $\frac{0.361/(1-0.361)}{0.577/(1-0.577)} = 0.414$. They observed that this simple adjustment greatly improved model calibration in the new sample.

**The problem**

Odds-ratio is a non-collapsible statistic: its estimate from the entire sample (marginal odds-ratio) is not equal to the average over subgroup-specific (conditional) values[4]. Because the intercept-adjustment approach applies an odds-ratio estimated from population-level quantities (marginal) to individual predictions (each individual being in its own subgroup), it is affected by non-collapsibility. To show this, consider a population consisting of equal number of low-risk and high-risk individuals with actual risks of, respectively, 1/10 and 1/2. Imagine we have an 'over-estimating' model that predicts risks of 1/7 and 3/5 for the two groups, respectively. For both groups, the odds of true risks is 2/3rd of the odds of predicted risks. Thus, applying an odds-ratio of 2/3 to predicted risks will resolve mis-calibration. However, using the intercept-adjustment method results in a different correcting odds-ratio: the overall average observed and predicted risks are 0.300 (odds of 0.429) and 0.371 (odds of 0.590), respectively, giving rise to marginal odds-ratio of 0.725. As this odds-ratio is closer to 1, the updated model under-corrects predicted risks, and this is a general pattern as the following section argues.

**Marginal odds-ratio under-corrects predictions**

Let $f(\pi, x)$ be the function that updates a predicted risk $\pi$ by applying an odds-ratio $x$:

$$\text{updated risk} = f(\pi, x) = \frac{\pi x}{1 - \pi + \pi x},$$

Let $p_0 = \mathrm{E}(\pi)$ and $p_1$ be the observed and desired average predicted risks, respectively. The goal of model updating is to solve $\mathrm{E}f(\pi, x) = p_1$ for $x$. But the marginal odds-ratio is a solution for $f(\mathrm{E}(\pi), x) = p_1$, and these two are not equal.

When $x < 1$, $f(\pi, x)$ is an increasing convex function of $x$, and by Jensen's inequality[5], $f(E(\pi), x) \leq Ef(\pi, x)$. Given this, the $x$ at which $f(E(\pi), x) = p_1$ will be higher than the $x$ at which $Ef(\pi, x) = p_1$. The reverse occurs for $x > 1$ where $f(\pi, x)$ is an increasing concave function. In both instances, the marginal odds-ratio under-corrects predicted risks. The marginal and conditional odds-ratios are equal only when there is no spread in predicted risks which does not hold for any useful prediction model.

**Improving the accuracy of intercept adjustment**

If a sample of predicted risks from the target population is available, model updating should be based on individual-level analysis[3]. The correct conditional odds-ratio is the exponent of the intercept term in a logistic model with the observed outcome as the dependent variable and the log-odds of predicted risks as the offset variable[6]. Janssen et al have proposed this method and the simple updating method as alternatives to each other; but this does not recognize the difference between marginal and conditional values[3].

If model updating needs to be performed based on summary statistics, one can still improve the accuracy of the simple method. By Taylor expansion of $f(\pi, x)$ for $\pi$ around its expected value, we have:

$$Ef(\pi, x) \approx f(E(\pi), x) + \frac{f''(E(\pi), x)}{2} var(\pi),$$

where $f''$ is the second derivative of $f()$ with respect to $\pi$. In this approximation, one should solve the following equation for $x$:

$$\frac{xp_0}{1-(1-x)p_0} + \frac{(1-x)x}{(1-(1-x)p_0)^3}v = p_1,$$

where $v$ is the variance of predicted risks.

This is a cubic equation whose closed form solutions can be obtained using standard methods. Re-arranging the terms and collecting coefficients for each power of x in the Taylor approximation equation will result in the following cubic equation:

$$(p_0^3 - p_1p_0^3)x^3 +$$
$$(3p_1p_0^3 - 2p_0^3 - 3p_1p_0^2 + 2p_0^2 - v)x^2 +$$
$$(p_0^3 - 3p_1p_0^3 + 6p_1p_0^2 - 2p_0^2 + p_0 - 3p_1p_0 + v)x +$$
$$(p_1p_0^3 - 3p_1p_0^2 + 3p_1p_0 - p_1) = 0$$

which can be solved in numerical (root-finding) or analytical methods. For example, the following R function, replying on the *RConics* package, will return the real roots based on closed-form solution for the cubic equation:

```
odds_adjust <- function (p0, p1, v)
{
  if (v > p0 * (1 - p0))
    stop("Variance cannot be larger than p0*(1-p0).")
  A <- p0^3 - p1 * p0^3
  B <- 3 * p1 * p0^3 - 2 * p0^3 - 3 * p1 * p0^2 + 2 * p0^2 -
    v
  C <- p0^3 - 3 * p1 * p0^3 + 6 * p1 * p0^2 - 2 * p0^2 + p0 -
    3 * p1 * p0 + v
  D <- p1 * p0^3 - 3 * p1 * p0^2 + 3 * p1 * p0 - p1
  res <- cubic(c(A, B, C, D))
  res <- Re(res[which(Im(res) == 0)])
  res <- res[which(res > 0)]
  res <- res[which(sign(log(res)) == sign(log(p1/p0)))]
  res
}
```

*Given the cubic nature of the associations, both the numerical root funding and the exact cubic polynomial solutions might return up to 3 real-valued ORs. In our investigations, the correct solution has always been obvious. The other ones were either negative, or on the*

> *wrong side of 1, e.g., if the target prevalence was lower than source prevalence, the odds-ratio was >1, or vice versa.*

**What is the impact of this correction?**

*Figure 1* demonstrates the results (in terms of relative bias) of brief simulations comparing the simple and Taylor methods. We modeled predicted risks as having beta distributions (often used to model probabilities). We considered target prevalence ($p_1$) that are 10%, 25%, and 50% lower (Panel A) or higher (panel B) than the original prevalence ($p_0$) under various values for $p_0$ and variance of the predicted risk.

*Figure 1:* Results of brief simulation studies comparing (Y-axis) the relative bias of the simple adjustment method (blue) and the Taylor method (green) against the exact conditional odds-ratio approach (red), as a function of the variance of predicted risks (main X-axis). The Area Under the Curve (AUC) of Receiver Operating Characteristic curve corresponding with each variance is also provided (second X-axis). Each panel pertains to a specific combination of original prevalence ($p_0$) and the desired change, in %, in calibration in the large (Delta)

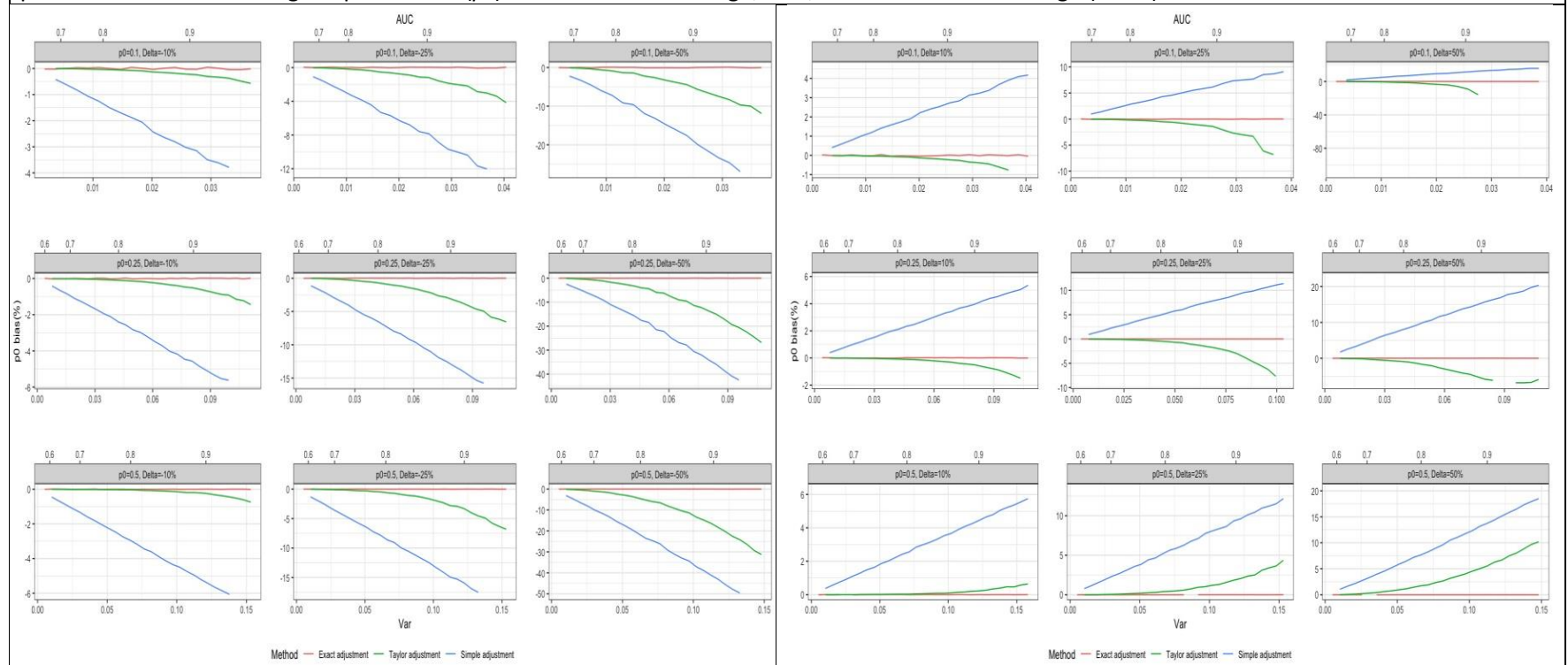

For each set, predictions are modeled as having beta(α,β) distribution, with values of α and β obtained from the mean ($p_0$) and variance using the method of moments. The exact method and AUC calculations were based on fitting a logistic regression model to $10^6$ randomly generated observations. Maximum variance is $p_0(1-p_0)$

Exact results are based on 10,000 simulated observations

AUC: Area under the Receiver Operating Characteristic curve

The Taylor method generally, and often substantially, reduced the bias of the simple method. The residual bias of this method was not large unless the Area Under the Receiver Operating Characteristic Curve (AUC, secondary X-axis) was very high (>0.9). The only scenario in which the Taylor method performed worse than the simple method was with $p_0 = 0.1$, $p_1 = 0.15$, and variance >0.026 (corresponding to AUC>0.93; top-right panel). In this scenario the distribution of predicted risks is bimodal and heavily right-skewed and thus the third moment is large. For variance values that resulted in AUCs ≤0.9, the Taylor method corrected ≥50% of the bias (often substantially more) of the simple method.

As a case study we consider ACCEPT, a model for predicting 12-month risk of exacerbations of Chronic Obstructive Pulmonary Disease, which provides predictions separately for all and severe exacerbations[7]. Since ACCEPT was developed among patients with a positive history of exacerbations, it overestimated the risk in patients with negative exacerbation history. We used patients with negative exacerbations history from the TORCH study[8] (n=571; 195 exacerbation [45 severe]) to correct the predictions. Ethics approval was obtained from The University of British Columbia Ethics Board (H11-00786). Starting from all exacerbations, the average observed risk in ACCEPT and TORCH were 0.56 and 0.34. The correct conditional odds-ratio based on applying the intercept-only logistic model was 0.358, whereas the marginal odds-ratio using the simple adjustment method was 0.403 (12.7% higher). The variance of predicted risks was 0.025, and thus the Taylor method resulted in a correcting odds-ratio of 0.362 (1.5% higher). For severe exacerbations, the average observed risks in development sample and TORCH were 0.183 and 0.079, respectively. The correct conditional odds-ratio was 0.338, whereas the marginal odds-ratio was 0.381 (12.6% higher). The variance of predicted risks was 0.024, and thus the Taylor method resulted in a correcting odds-ratio of 0.337 (0.2% higher).

**Remarks**

The non-collapsibility of odds-ratio is a widely discussed phenomenon in the causal inference literature, cited as the cause of misinterpretations and non-intuitive observations such as the classical Simpson's paradox[9]. In predictive analytics, the numerical value of effect size is not of primary interest and coefficients are a 'means to an end' which is to make good predictions. However, non-collapsibility is first and foremost a mathematical property of a measure, not its interpretation, and, as we showed, can have ramifications in predictive analytics. We showed that a guideline-recommended[10] model updating method under-corrects predictions. The degree of under-correction grows with the variance of predicted risks, thus particularly affecting models with high discrimination as they generate predictions with high dispersion.

To mitigate such under-correction, we proposed an alternative adjustment method based on Taylor approximation of the conditional odds-ratio. This method requires an estimate of predicted risks' variance. This is an important metric that risk modeling studies should ideally report. If not, one can often find indirect evidence for it. For example, the inter-quartile range or the histogram of predicted risks can be used to approximate variance. Many studies report calibration plots that on their X-axis show the dispersion of predicted risks, which can be used to infer about both mean and variance (with histogram or calibration plots one can quantify other moments and include more terms in Taylor approximation to further improve the adjustment).

Going back to the post-operative pain example, Janssen et al did not report the variance of predicted risks, but by digitizing the calibration plot and reading the decile values on the X-axis, we estimated $var(\pi) \sim 0.025$. This will result in an odds-ratio of 0.375 (vs. 0.414 from simple intercept-adjustment). Even if such indirect evidence is unavailable, making an educated guess about (e.g., based on similar prediction models) or a Bayesian approach that assigns a probability distribution to variance is likely a better substitute than a method that is guaranteed to under-correct the predictions. The R code implementing the Taylor method is provided as part of the *predtools* package (https://github.com/resplab/predtools).